\documentclass[12pt]{article}
\usepackage[a4paper]{geometry}
\usepackage{graphicx}
\usepackage{amssymb}
\usepackage{epstopdf}
\usepackage[applemac]{inputenc}
\usepackage{times}
\usepackage{url}
\usepackage{setspace}
\usepackage{lineno}
\usepackage{cite}

\graphicspath{{figures/}}

\providecommand{\figref}[1]{\ref{fig:#1}}
\providecommand{\fig}{Fig.}
\providecommand{\figr}[1]{\fig\,\figref{#1}}
\providecommand{\figlabel}[1]{\label{fig:#1}}

\title{Böögg Bang drives global climate change}
\author{M.~S.~Brennwald, D.~M.~Livingstone and R.~Kipfer\\
{\small Eawag, Swiss Federal Institute of Aquatic Science and Technology, CH-8600 Dübendorf, Switzerland}\\
{\small matthias.brennwald@eawag.ch}
}

\date{\small April 1, 2011}

\begin{document}
\maketitle


\begin{abstract}
The Böögg is a large model of a snowman, constructed of inflammable materials and filled with explosives. During the traditional festival of Sechseläuten, which takes place each spring in Zurich, Switzerland, the Böögg is placed atop a wooden pyre, which is set alight. According to popular legend, the time that elapses until the Böögg's head explodes (the "head-bang" time) is said to give a rough forecast of local weather conditions prevailing during the following summer. However, recent research has questioned the validity of this prediction. To study the Böögg's predictive powers, we analyzed the Böögg head-bang time record from 1965-2010 within the context of global climate change. Our analysis shows that the Böögg head-bang time is a good predictor not of short-term local weather, as might be expected from the legend, but of the behavior of the entire global climate system.
\end{abstract}

\section{Introduction}
Sechseläuten is a famous and prestigious climate conference held each spring in Zurich, Switzerland. Sechseläuten can look back on a much longer history than most scientific conferences, and in fact in its original form it dates back to medieval times. To boost the number of participants, the conference is traditionally open to the public\footnote{With the exception of women, who were sometimes allowed as active participants at Sechseläuten during medieval times. During the last few centuries and until 2010, however, they were tolerated only as passive spectators.} and is even disguised as a public holiday and promoted as a tourist attraction \cite{boeoegg_encyclopedia} in a rare example of integrated transdisciplinary multiple socio-functionality (which other scientific conferences would do well to emulate). The climax of Sechseläuten is reached during a practical workshop held as an integral part of the conference, in which the ritual burning of the so-called ``Böögg'' (\figr{boeoegg_photo}), a strange, pale, humanoid doll not unlike a snowman in appearance, takes place. Interestingly, the word ``Böögg'' is related linguistically to the everyday English words ``bogey'' (snot),``bogeyman'' (child-eating monster), ``bugger'' (sodomite), and ``Bulgarian'' (someone from Bulgaria). The Böögg differs from most normal snowmen in that it is flammable and that its body, and most particularly its head, are filled with explosives. Prior to the workshop, the Böögg is set by the organizers (the Zentralkomitee der Zünfte Zürich, ZZZ) on a ritual funeral pyre consisting of a pile of wood $\sim$10 m in height. During the workshop, the pyre is ignited, and the time that elapses between initial ignition of the pyre and explosion of the Böögg's head (known as the ``head-bang time'') is noted. Each year, in contrast to most scientific conferences, the Sechseläuten conference produces a concrete result; viz.\ a 6-month forward regional weather prediction, based on the head-bang time. A short head-bang time presages a warm, sunny summer, and a long head-bang time a cold, wet summer. Recent work, however, has questioned the validity of the Böögg prediction \cite{Arnet:2005,Primault:2005,Schmuki:2006}.

\begin{figure}[tbp]
   \centering
   \mbox{
   \includegraphics[width=0.32\linewidth]{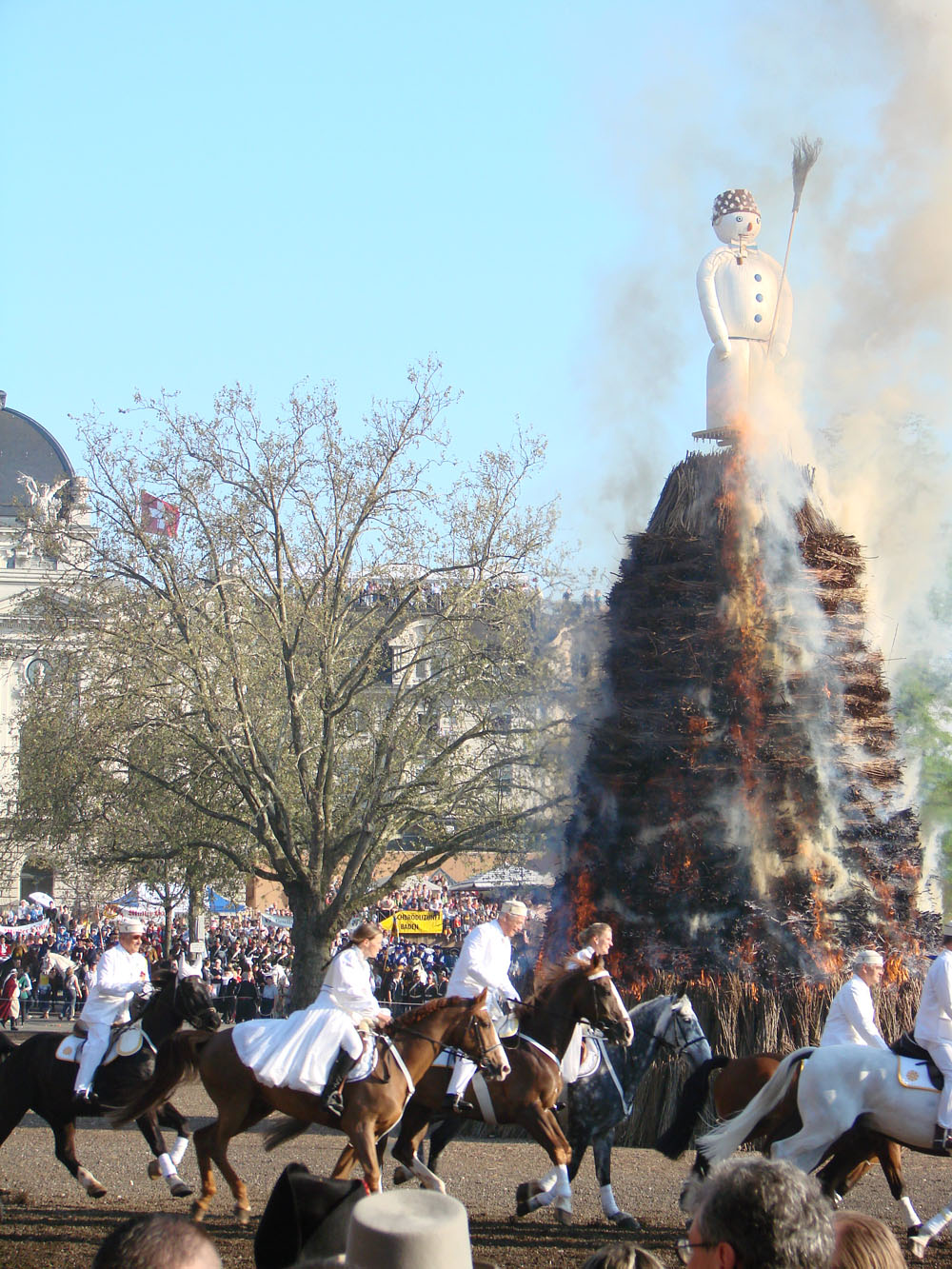}
   \includegraphics[width=0.32 \linewidth]{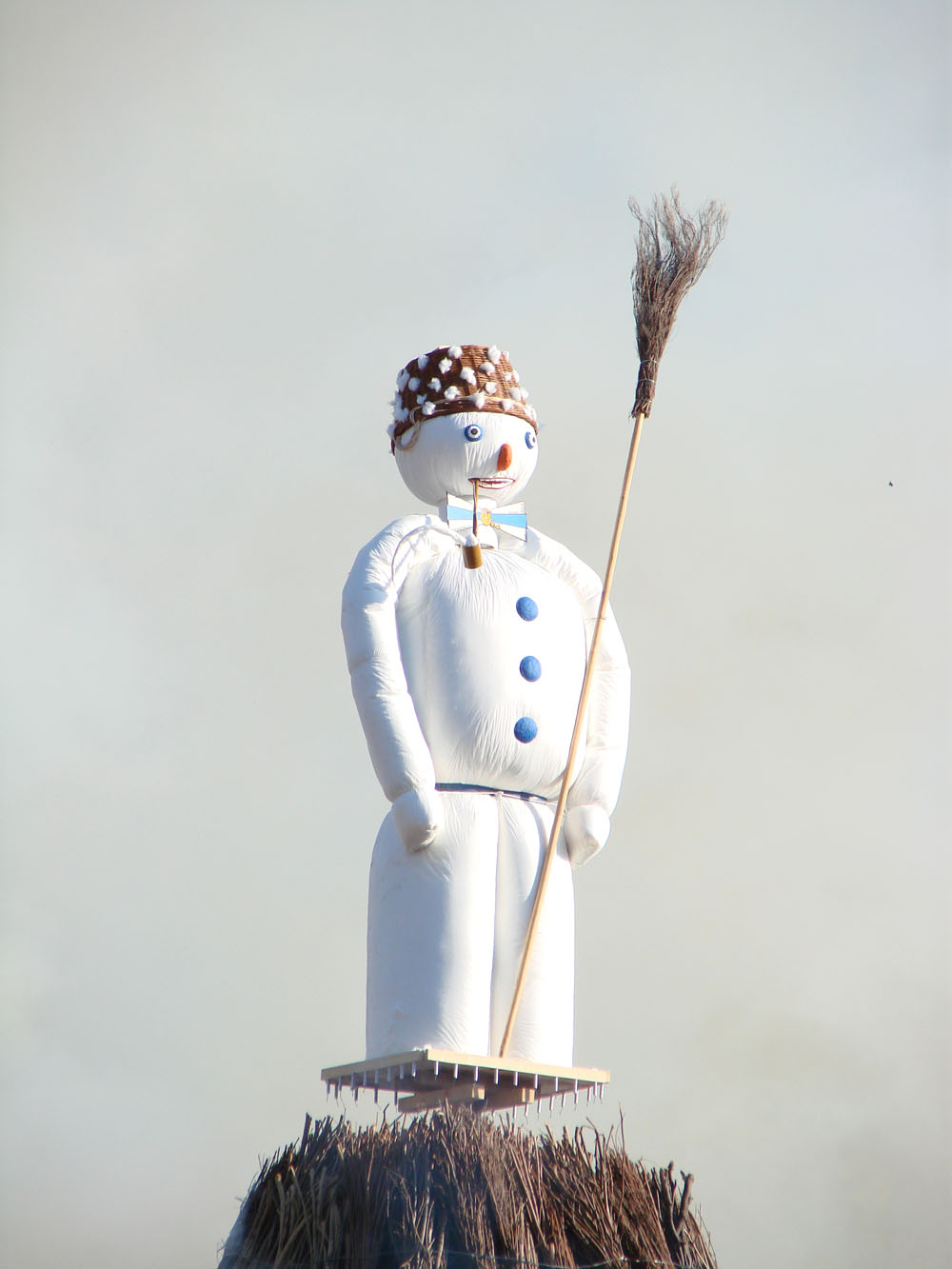}
   \includegraphics[width=0.32 \linewidth]{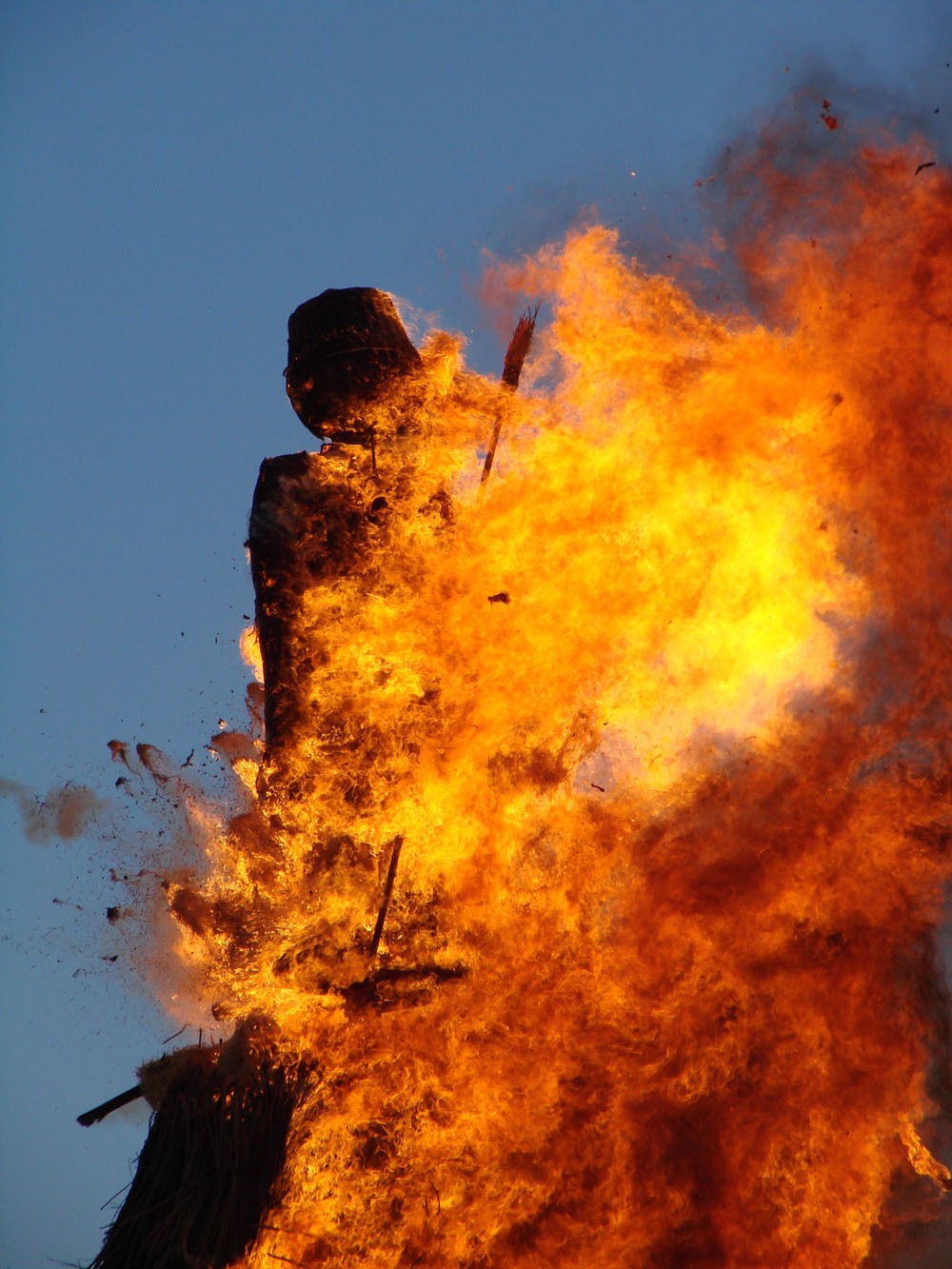}
   }
   \caption{Left: Böögg and Sechseläuten organizers (under hats or on horses). Center: close-up of the Böögg. Right: explosion of the Böögg.
  }
   \figlabel{boeoegg_photo}
\end{figure}

To assess whether the quality, fame and reputation of the Sechseläuten conference for producing scientifically accurate predictions of weather and climate conditions is justified, we conducted a reanalysis of the Böögg head-bang times for 1965--2010. Furthermore, we evaluated this unique record not merely with regard to its short-term predictions of weather and climate conditions in the Zurich region (which is of special interest only to the operators of Swiss outdoor swimming baths and ice-cream parlors), but within the much broader and more interesting context of global climate change.\par

Climate change has been the subject of much recent scientific attention, not only at the Sechseläuten conference but within the scientific community in general. One notable aspect of climate change is that climate conditions often change abruptly rather than evolving gradually. Two of the most prominent examples of such a climate regime shift are those that occurred almost simultaneously all around the world in the mid-1970s and the late 1980s \cite{Hare:2000}, affecting physical and biological conditions in the oceans \cite{Reid:2001,Alheit:2005,Rodionov:2005a,Zhang:2007,Tian:2008,Conversi:2009,Conversi:2010,Kidokoro:2010}, lakes \cite{Arpe:2000,Gerten:2000,Peeters:2000a,Anneville:2004,Anneville:2005,Temnerud:2008}, rivers \cite{Hari:2006}, and even groundwater \cite{Figura:2009}. The late 1980s climate regime shift also gave rise to an abrupt decrease in the number of snow days in the Swiss Alps \cite{Marty:2008}, severely affecting winter tourism.\footnote{And thus potentially affecting the number of participants at the Sechseläuten climate conference. The ramifications of this positive feedback loop for global climate change, and, more importantly, for Swiss winter tourism, is currently the subject of a transdisciplinary study being conducted jointly with the Swiss tourist board.}\par

Given the importance of abrupt climate regime shifts within the current climate change debate, we decided to investigate statistically the capability of the Böögg head-bang time to predict empirically the occurrence and timing of climate regime shifts.\par

\section{Analysis of the Böögg head-bang time-series}
We analyzed the time-series of Böögg head-bang times from 1965--2010 (\figr{boeoegg_data}). Data from 1965--2004 were taken from an earlier assessment of the Böögg's forecasting abilities of weather and climate conditions \cite{Arnet:2005}. The data from 2005--2010 were obtained from the website of a prominent Swiss tabloid newspaper \cite{blick_webpage}, because of the superiority of its search engine over those of more elitist broadsheet newspapers. \footnote{Unrelated side note: a well known scientific ``tabloid'' journal reported \cite{fame-nature} on a notable link between scientific fame, a popular search engine and Janet Jackson \cite{Bagrow:2004}.} 
Data prior to 1965 were not utilized, as the ZZZ's proceedings of the Sechseläuten conference were either inaccessible, incomplete, ambiguous or illegible and therefore did not meet the standard of quality required for our analysis.\par

\begin{figure}[tbp]
   \centering
   \includegraphics[width=\linewidth]{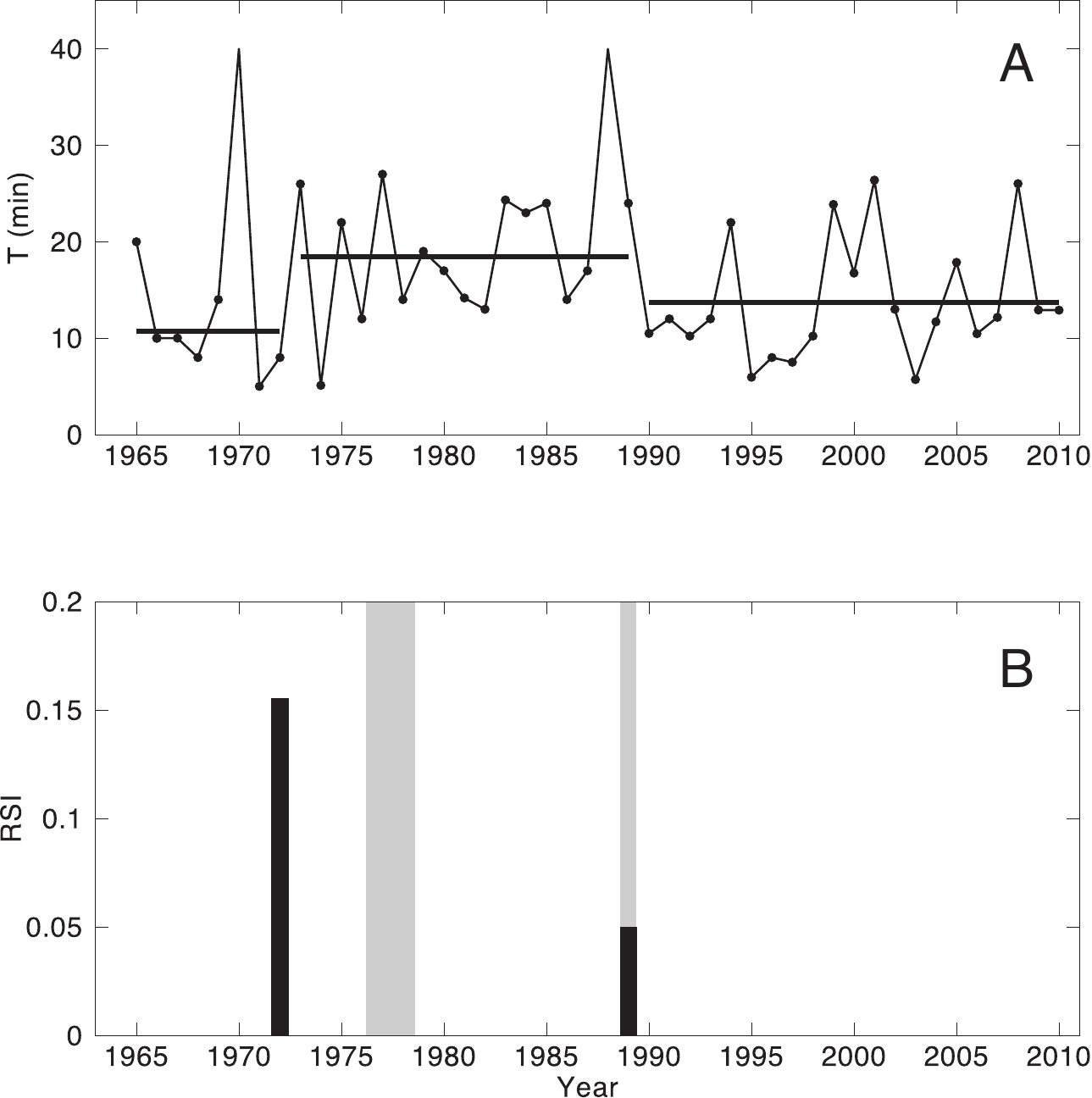} 
   \caption{A: Head-bang time (T) of the Zurich Sechseläuten Böög since 1965. Horizontal lines indicate mean head-bang times during the periods 1965--1972, 1973--1989, and 1990--2010. Data values from 1970 and 1988 were considered outliers and therefore excluded from the analysis. B: Regime Shift Index (RSI) for the head-bang time, calculated using the Rodionov regime-shift detection method \cite{Rodionov:2004} ($p$-value\,=\,0.05, Rodionov cut-off length $l$\,=\,10). Gray shaded areas represent large-scale climate regime shifts reported in the literature.}
   \figlabel{boeoegg_data}
\end{figure}

To identify climate regime shifts in the Böögg head-bang time-series, we used an algorithm that was specially designed to detect climate regime shifts \cite{Rodionov:2004,Rodionov:2005a}. The algorithm yields a regime shift index (RSI) for those data values that are suspected to be located at a regime shift. Although the result is rather sensitive to the statistical parameters used (viz.\ the $p$-value of the regime shift and the cut-off length $l$ corresponding to the maximum length of the data sequences to be considered by the algorithm), using expert scientific judgement\footnote{We used the values everyone else uses.} we were able to obtain the RSI values illustrated in \figr{boeoegg_data}.

\section{Discussion, conclusions, and speculations}
The regime-shift pattern of the Böög head-bang time is strikingly similar to the pattern of large-scale climate regime shifts observed in the mid-1970s and in the late 1980s (\figr{boeoegg_data}), implying that the Böögg head-bang time and global climate change are tightly related.\par

It is well known that the application of statistical methods to data is similar to the application of cooking recipes to raw food ingredients, where the final dish obtained depends not only on the recipe, but also on the cooks. Despite this fact, and being fully aware of it, we are convinced that the relationship between the Böög head-bang time and global climate change apparent from \figr{boeoegg_data} represents more than mere coincidence. A fundamental question related to cause and effect therefore arises: Is the Böögg head-bang time being driven by global climate, or is global climate being driven by the Böögg head-bang time? Related to this is a further question: What are the mechanisms relating the two? We believe we can answer both these questions satisfactorily.\par

The mid-1970s climate regime shift (localized approximately in 1976/77) follows the first of the two regime shifts observed in the Böögg head-bang time-series (the RSI peak in 1972) by several years; it is therefore impossible that the Böögg head-bang time is being driven by global climate. Following S.~Holmes: ``When you have eliminated all which is impossible, then whatever remains, however improbable, must be the truth'' \cite{Doyle:1927}. Therefore, global climate is being driven by the Böögg head-bang time. Although it may not be intuitively obvious to many readers how the explosion of a snowman's head in Zurich can drive global climate, it should be borne in mind that the gnomes of Zurich (the local bankers \cite{BBC}) are genetically close relatives of the Zurich Böögg, and that their main underground cave system is located only a few hundred meters from the Böögg's annual funeral pyre. Some of the more sociable gnomes even participate in the Sechseläuten conference. Information on the Böögg's head-bang time and its apparent climatic implications is therefore immediately available to the gnomes, who then adjust their financial and economic activities, insofar as these are climatically dependent, to correspond to the current Böögg predictions. They consider the Böög predictions to be superior to the predictions of global climate models, which they automatically assume to have inferior predictive powers similar to those of economic forecasting models, with which they are most familiar. \figr{boeoegg_data} provides support for this hypothesis. The lag time between the 1972 RSI peak in the Böögg head-bang time-series and the mid-1970s climate regime shift reflects the time taken for the financial activities of the gnomes of Zurich to spread into the global economy, and then to affect global climate. As a result of the increasing financial and economic globalisation that occurred in the 1970s and 1980s, coupled with simultaneous innovations in communications technology and the introduction of business strategies resulting in ``just-in-time'' inventory and manufacturing practices, by the late 1980s this lag time had decreased to the point at which the impact of the gnomes' activities on the global economy had become effectively instantaneous. Thus by the late 1980s the impact of the explosion of the Böögg's head on global climate had been reduced to a few months at most. The late 1980s head-bang time regime shift was thus reflected extremely rapidly in global climate.\par

To explore further our hypothesis about the nature of the causal chain linking the Böögg head-bang time to global climate change, better knowledge of the genetic relationship between the Zurich Böögg and the gnomes of Zurich would be required. We expect that recent extensions to the Human Genome Project \cite{Wu:1991} will be useful in providing further insight into the mechanisms linking global climate change to the behavior of the Böögg.\par

Even although we were not able to elucidate fully all the details of the mechanisms relating global climate change to the Böögg head-bang time, our findings confirm that the Böögg head-bang time does indeed provide an accurate prediction of future climate conditions. However, in contrast to the traditional view propagated by the Sechseläuten climate conference organizers, the Böögg head-bang time prediction is valid not only for the spatially and temporally limited case of local summer weather conditions in the Zurich region; instead, the Böögg head-bang time governs the long-term behavior of the entire global climate system.\par

\bibliographystyle{boeoegg}
\bibliography{boeoegg_extralit,MB_Literatur}

\end{document}